\documentclass[twocolumn]{aastex63}
\usepackage{amsmath}
\usepackage{hyperref}
\usepackage{datetime}
\usepackage{appendix}
\usepackage{comment}
\usepackage{chngcntr}
\usepackage{soul}
\usepackage{bm}
\usepackage{xcolor}

\def\msun{h^{-1}M_{\odot}}
\def\mpc{h^{-1} {\rm Mpc}}
\def\kpc{h^{-1} {\rm kpc}}
\def\hmpci{h{\rm Mpc}^{-1}}

\newcommand{\avrg}[1]{\left\langle #1 \right\rangle}

\begin{document}

\begin{flushright}
YITP-21-42
\end{flushright}

\title{An Optimal Estimator of Intrinsic Alignments for Star-forming Galaxies in IllustrisTNG Simulation}

\author[0000-0001-9879-4926]{Jingjing Shi}
\affiliation{Kavli Institute for the Physics and Mathematics of the Universe (WPI), The University of Tokyo Institutes for Advanced Study (UTIAS),
The University of Tokyo, 5-1-5 Kashiwanoha, Kashiwa-shi, Chiba, 277-8583, Japan}

\author[0000-0002-7934-2569]{Ken~Osato}
\affiliation{Institut d'Astrophysique de Paris, Sorbonne Universit\'{e}, CNRS, UMR 7095, 75014 Paris, France}
\affiliation{Center for Gravitational Physics, Yukawa Institute for Theoretical Physics,
Kyoto University, Kyoto 606-8502, Japan}
\affiliation{LPENS, D\'epartement de Physique, \'Ecole Normale Sup\'erieure,
Universit\'e PSL, CNRS, Sorbonne Universit\'e, Universit\'e de Paris, 75005 Paris, France}

\author{Toshiki~Kurita}
\affiliation{Department of Physics, The University of Tokyo, 7-3-1 Hongo, Bunkyo-ku, Tokyo 113-0033 Japan}
\affiliation{Kavli Institute for the Physics and Mathematics of the Universe (WPI), The University of Tokyo Institutes for Advanced Study (UTIAS),
The University of Tokyo, 5-1-5 Kashiwanoha, Kashiwa-shi, Chiba, 277-8583, Japan}

\author[0000-0002-5578-6472]{Masahiro~Takada}
\affiliation{Kavli Institute for the Physics and Mathematics of the Universe (WPI), The University of Tokyo Institutes for Advanced Study (UTIAS),
The University of Tokyo, 5-1-5 Kashiwanoha, Kashiwa-shi, Chiba, 277-8583, Japan}

\email{jingjing.shi@ipmu.jp}

\begin{abstract}
Emission line galaxies (ELGs), more generally star-forming galaxies, are valuable tracers of large-scale structure and therefore main targets of upcoming wide-area spectroscopic galaxy surveys. We propose a fixed-aperture shape estimator of each ELG for extracting the intrinsic alignment (IA) signal, and assess the performance of the method using image simulations of ELGs generated from the IllustrisTNG simulation including observational effects such as the sky background noise. We show that our method enables a significant detection of the IA power spectrum
with the linear-scale coefficient $A_{\rm IA}\simeq (13$--$15)\pm 3.0$ up to $z=2$, even from the small simulation volume $\sim0.009\,(h^{-1}{\rm Gpc})^3$, in contrast to the null detection with the standard method. Thus the ELG IA signal, measured with our method, opens up opportunities to exploit cosmology and galaxy physics in high-redshift universe.
\end{abstract}

\section{Introduction}
\label{sec_intro}

Intrinsic correlations between shapes of different galaxies, the so-called intrinsic alignments (IA), can arise from the primordial seed fluctuations of cosmic structures and gravitational interaction in structure formation \citep{2000ApJ...532L...5L,2001ApJ...555..106L,2000ApJ...545..561C,2000ApJ...543L.107P,Catelanetal2001,2001ApJ...559..552C}. IA has been intensively studied because it is one of the most important systematic effects for weak lensing
\citep{HirataSeljak2004,2015PhR...558....1T} and can also be used as a probe of cosmological parameters \citep{2013JCAP...12..029C,2020ApJ...891L..42T},
the primordial gravitational wave \citep{2014PhRvD..89h3507S,2015JCAP...10..032S}, and the primordial anisotropic non-Gaussianity \citep{2016PhRvD..94l3507C,2020arXiv200703670A}.

Observationally, several studies have shown a significant detection of the IA effect for luminous red galaxies up to $z\sim 0.7$ \citep{2006MNRAS.367..611M,2007MNRAS.381.1197H,2009ApJ...694L..83O,2015MNRAS.450.2195S,2016MNRAS.457.2301S}.
In contrast, there is no clear signature of the IA effect reported for blue star-forming galaxies from data \citep{2011MNRAS.410..844M,2018PASJ...70...41T,2020ApJ...904..135Y}; and different simulations predict IA with varying amplitudes and signs \citep{2017MNRAS.468..790H,2016MNRAS.461.2702C,2016MNRAS.462.2668T,2021JCAP...03..030S}. 
Star-forming galaxies are useful tracers of large-scale structures up to the higher redshifts, and indeed main targets for upcoming wide-area galaxy redshift surveys such as the Subaru Prime Focus Spectrograph (PFS) survey \citep{2014PASJ...66R...1T} and
the DESI survey\footnote{\url{https://www.desi.lbl.gov}}.

Hence the purpose of this paper is to study a new estimator of shapes of star-forming galaxies enabling a significant detection of their IA effect. To do this we use star-forming galaxies simulated in the IllustrisTNG \citep{Springel2018}. We use our method to characterize shapes of the galaxies, and then measure the IA power spectrum based on the method in \citet{2021MNRAS.501..833K} \citep[also see][]{2021JCAP...03..030S}. 
To make the realistic predictions, we take into account observational effects (sky background, filter transmission and seeing) when characterizing the galaxy shapes. 

The structure of this paper is organized as follows. In Section~\ref{sec_method} we briefly review the IllustrisTNG simulation and describe our selection of ELGs from the simulation. 
In Section~\ref{sec:method} we describe a method to simulate observed images of ELGs with a ground-based 
telescope, and then define a new estimator of ELG shapes. 
In Section~\ref{sec_res} we show the main results of this paper: the IA power spectra measured using the 
new estimator of ELG shapes. Section~\ref{sec_con} is devoted to conclusion and discussion.
Throughout this paper we use the comoving coordinates to refer length scales. 

\section{Simulation data and ELG selection}
\label{sec_method}

\subsection{The IllustrisTNG simulation suite}
\label{sec:illustris}

The IllustrisTNG is a suite of cosmological hydrodynamical simulations in a $\Lambda$CDM model
\citep[][also see references therein]{Springel2018,Marinacci2018,Naiman2018,Pillepich2018,Nelson2018}.  
In this work, we use the publicly-available simulation data of
IllustrisTNG-300 (hereafter simply TNG300) 
\citep{2019ComAC...6....2N} with a box size of about $205\mpc$ to have good statistics.
The simulation assumes the \textit{Planck} cosmology \citep{2016A&A...594A..13P}, characterized by
$\Omega_{\rm m}=0.3089$, $\Omega_{\rm b}=0.0486$, $h=0.6774$, and $\sigma_{\rm 8}=0.8159$. It follows the dynamical evolution of $2500^3$ dark matter (DM) particles and approximately $2500^3$ gas cells from $z=127$ to $z=0$, giving an average gas cell mass of $7.44\times 10^6\msun$, a DM mass resolution of $3.98\times 10^7\msun$, and a collision-less softening length of $1\kpc$ at $z=0$. The TNG300 galaxy formation and evolution model includes key physical effects such as gas cooling and heating, star formation, stellar evolution and chemical enrichment, SN feedback, BH growth, AGN feedback, and cosmic magnetic field (see \citealt{Weinberger2017, Pillepich2018model} for details).

\subsection{ELG selection}
\label{sec:ELG_selection}

Galaxies in TNG300 are identified using the {\tt Friends-of-Friends} (FOF) and {\tt SUBFIND} algorithms \citep{Davis1985,Springel2001}. We select hypothetical emission-line galaxies (ELGs) from a ranked list of the star formation rate (SFR) of galaxies until their comoving number density matches $n_{\rm g}=10^{-4}({\mpc})^{-3}$, a typical number density of ELGs that upcoming surveys such as PFS and DESI are designed to have. 
The SFR ranked selection roughly corresponds to a selection of ELGs based on the [{\sc O\,ii}] emission line strengths (\citealt{2020MNRAS.498.1852G,Osato2021}). Here the SFR of each galaxy in the simulation is defined by the {\it spontaneous} SFR within the twice half-stellar-mass radius. In this work, we focus mainly on ELGs at $z=1.5$, as a representative redshift of PFS and DESI surveys. We will also use ELGs at different redshifts $z=0.5$, $1$, and $2$, selected in the same way, to study the redshift dependence of their IA signals.
Table~\ref{tab_elg} shows properties of ELGs used in our study. More than $67\%$ of the ELGs reside in central subhalos and are massive in the stellar mass.

\begin{table}[h]
\centering
\scriptsize
\caption{Properties of ELGs in Illustris-TNG300, studied in this work. We show the redshift ($z$), average stellar mass ($M_\star$), average host halo mass ($M_{\rm 200}$), average star formation rate, the central galaxy fraction, IA strength ($A_{\rm IA}$), and the rms ellipticities per component ($\sigma_{\epsilon}$). Here  $M_\star$ and $M_{\rm halo}$ are in units of $\msun$, and SFR is in units of $M_{\odot}\,{\rm yr}^{-1}$.}
\begin{tabular}{l|c|c|c|c|c|c} \hline\hline 
$z$ & $\avrg{{\rm log} M_\star}$ & $\avrg{{\rm log} M_{\rm halo}}$ & $\avrg{{\rm SFR}}$ & $f_{\rm cen}$ & $A_{\rm IA}$ & $\sigma_\epsilon$ \\
\hline 
0.5 & 10.39 & 13.20 & 25.75 & 0.667 & $15.39\pm 2.96$ & 0.43\\
1.0 & 10.41 & 13.04 & 47.78 & 0.682 & $15.26\pm 2.89$ & 0.41\\
1.5 & 10.42 & 12.88 & 71.64 & 0.741 & $12.86\pm 2.83$ & 0.39\\
2.0 & 10.41 & 12.67 & 94.01 & 0.798 & $15.45\pm 2.84$ & 0.40\\
\hline
\hline
\end{tabular}
\label{tab_elg}
\end{table}
%

\section{Method: image simulation and aperture-based shape estimator}
\label{sec:method}

\subsection{An image simulation of ELGs}
\label{sec_simulated_image}

\begin{figure*}
\centering
 \includegraphics[width=1.\linewidth]{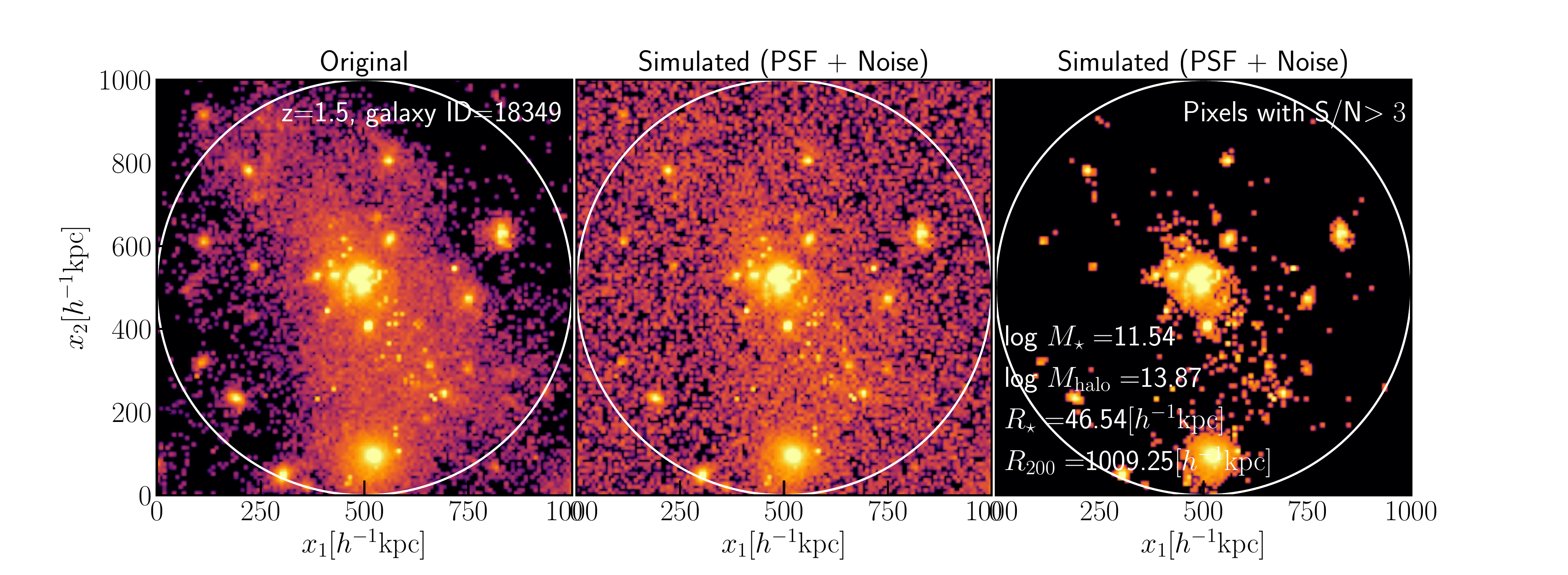}
\caption{\label{fig_gal_image} A simulated $i$-band image of the region around an example ELG at $z=1.5$, made from the TNG300 simulation data. 
{\it Left panel}: The original image of the ELG region. 
{\it Middle}: The simulated image taking into account the $0.6^{\prime\prime}$ FWHM seeing effect
, the total system throughput of the Subaru telescope ($0.5$), and the sky background noise at the Subaru site, assuming $t_{\rm exp}=1,200\,{\rm sec}$ for the exposure time and the $i$-band filter transmission.
{\it Right}: Similar to the middle panel, but it shows only the pixels with ${S/N}>3$ within an aperture of radius $500\kpc$ around the ELG. The legend gives the stellar mass and the half-stellar-mass radius of the ELG, and the halo mass and the virial radius of the host halo.}
\end{figure*}

In the following we assume that both imaging and spectroscopic data for ELGs in the sample are available; 
a spectroscopic redshift of each ELG, via an identification of the emission-line(s) such as 
[{\sc O\,ii}] line, and an imaging data around each ELG, including surrounding galaxies that do not necessarily have spectroscopic redshifts.
Here the spectroscopic redshifts are needed to measure the 3D power spectrum of ELG's IA effect, and the images are needed to characterize shapes of the regions surrounding ELGs for the IA measurements.  
Such imaging and spectroscopic surveys for the same region of the sky are available for ongoing and upcoming surveys such as the Subaru Hyper Suprime-Cam (HSC) and PFS surveys. 
We below describe a method to simulate an imaging data of each ELG region that is seen with a ground-based telescope such as the Subaru HSC.

We first carry out a ray-tracing simulation of each ELG region in TNG300 to obtain the projected image. 
We use all the stellar particles contained in a cubic box of $(1\mpc)^3$ volume around each ELG at the center.
The spectral energy distributions (SEDs) of the stellar population are modeled with the stellar population synthesis code {\tt P\'EGASE.3} \citep{2019A&A...623A.143F}.
Each stellar particle represents a single age stellar population.
First, we construct the table of SEDs for different metallicities $Z = [0.0, 0.0001, 0.0004, 0.004, 0.008, 0.02, 0.05, 0.1]$
up to the age of $100 \, \mathrm{Myr}$.
Then, for each particle, we allocate SEDs by linearly interpolating the table with respect to the metallicity and age.
The attenuation due to diffuse interstellar medium and dust is taken into account in {\tt P\'EGASE.3.}
Using the {\emph{rest-frame}} luminosity per unit wavelength and the luminosity distance to the galaxy redshift (e.g. $z=1.5$), we calculate the {\emph{observer-frame}} flux per unit wavelength.
Then we include the filter transmission to calculate
the noise-free and PSF-free photon counts in each pixel of the simulated image, taking the $x_3$-direction as the line-of-sight direction, as shown in the left panel of Fig.~\ref{fig_gal_image}. 
In doing these we assume the atmosphere transparency of 1.0, the aperture of the 8.2m Subaru Telescope, the total system throughput of $0.5$, 
$t_{\rm exp}=1,200\,{\rm sec}$ for the exposure time, and the transmission of $i$-band, more exactly the
$i2$-filter of Subaru HSC\footnote{\url{https://www.subarutelescope.org/Observing/Instruments/HSC/sensitivity.html}} that has a transmission curve over $689<\lambda/[{\rm nm}]<845$.
We generate a simulated image of each ELG in $128^2$ pixels for a square region of $1\,(h^{-1}{\rm Mpc})^2$ around the ELG.
The pixel size is $7.8\,h^{-1}{\rm kpc}$ corresponding to 0.53~{\rm arcsec} for a galaxy at $z=1.5$.

We then include the atmospheric effects.
The turbulence of the atmosphere smears
the image resolution -- the seeing effect. 
To model the seeing effect, we convolve the above {\emph{observer-frame}} image with a 2D Gaussian function with FWHM$=0.6$\,arcsec, which is a typical seeing size of the HSC data \citep{2018PASJ...70S...4A}. 
In addition, the sky itself emits light -- the sky background. Assuming the sky background dominated regime, we generate the random noise in each pixel, assuming a Gaussian distribution with width $\sigma_{\rm sky}=2849 e^{-}/{\rm s}/{\rm arcsec}^2$ (electron counts per second per ${\rm arcsec}^2$ solid angle) and 
$t_{\rm exp}=1,200\,{\rm sec}$ for the exposure time, where $\sigma_{\rm sky}$ is obtained from the HSC ETC\footnote{\url{https://hscq.naoj.hawaii.edu/cgi-bin/HSC_ETC/hsc_etc.cgi}}
assuming an observation at $7$ days after new moon with moon-object distance of $90$ degrees. 
Our simulated image fairly well reproduces $i_{\rm lim}\simeq 25.7$ for the $5\sigma$ 
limiting magnitude ($2^{\prime\prime}$ aperture) for a point source as obtained in the HSC ETC, and this depth is roughly equivalent to the depth of the ongoing 
Subaru HSC survey \citep{2018PASJ...70S...4A}.

Fig.~\ref{fig_gal_image} shows the simulated image in the region around an example ELG. 
This ELG resides at the central subhalo, and the host halo has the virial radius $R_{\rm 200}\simeq 1\,h^{-1}{\rm Mpc}$ 
($M_{200}\simeq 7.4\times 10^{13}\,h^{-1}M_\odot$), greater than the panel size, whilst the ELG itself has a half-stellar-mass radius of $R_\ast\simeq 46~{\kpc}$, much smaller than $R_{\rm 200}$. 
The figure shows that the ELG is surrounded by satellites or many building blocks, which would accrete onto the ELG to form a bigger galaxy at lower redshifts. The accretion direction should reflect shapes of the host halo and surrounding cosmic web, and the method we propose below is sensitive to these building blocks to better capture the overall IA signal. However, some of these building blocks become invisible when the sky noise is added,  
as shown in the middle panel. The right panel shows the pixels that have $S/N\ge 3$, and bright building blocks survive even after the $S/N$ cut.

\subsection{An aperture shape estimator for ELGs}
\label{sec:inertia_def}

We now characterize the ``shape'' of each ELG using the simulated images around each ELG we described in the preceding section.
In this work, we propose an ``aperture inertia tensor'' for ELG shapes, defined as
\begin{equation}
    I^{\rm ap}_{ij}=\frac{\sum_{n; (S/N)_{\rm pix}>3; r^{\rm 2D}_n\le 500{\kpc}}~ f_n x_{ni} x_{nj}}{\sum_{n; (S/N)_{\rm pix}>3;r^{\rm 2D}_n\le 500{\kpc}}~ f_n},
\label{eq:aperture_Iij}
\end{equation}
where $f_n$ is the flux of the $n$-th pixel in the simulated image, $x_{ni}$, $x_{nj} (i,j=1,2)$ are the relative position of this pixel with respect to the ELG position, 
and the summation runs over all the pixels within circular aperture of the projected radius $r^{\rm 2D}\le 500\kpc$ that have $S/N\ge 3$ for the signal-to-noise ratio of photon counts in the pixel. 
The average virial radius of the host halos for ERGs at $z=1.5$ is $\avrg{R_{200}}\simeq 470\,h^{-1}{\rm kpc}$, corresponding to 
the average halo mass $\avrg{M_{200}}\simeq 7.6\times 10^{12}\,h^{-1}M_\odot$ (Table~\ref{tab_elg}), and roughly matches the aperture radius. 
Note that we use the fixed aperture of $r^{\rm 2D}_{\rm ap}=500\,h^{-1}{\rm kpc}$ for all the results in this paper. We also test our results with smaller aperture sizes, such as $200\kpc$ or $300\kpc$, as shown in Appendix~\ref{sec_aperture_size}.

We find that the inertia tensor is ill-defined if we do not employ the $S/N$ cut. However, the results basically do not change if we adopt different $S/N$ cuts such as $S/N>4$ or $S/N>5$. For the above inertia tensor, stellar particles at outer radii are up-weighted so that the estimator can capture contribution from building blocks around each ELG as seen in the middle and right panels of Fig.~\ref{fig_gal_image}.

For comparison, we also study the conventionally used inertia tensor for the same sample of ELGs.
The reduced inertia tensor is widely used \citep{2015MNRAS.448.3522T}, 
\begin{equation}
    I^{\rm reduced}_{ij}=\frac{\sum_{n} m_n \frac{x_{ni} x_{nj}}{r_n^2}}{\sum_{n} m_n},
\label{eq:reduced_Iij}
\end{equation}
where $m_n$ is the mass of the $n$-th {\it member} stellar particle of the ELG, $x_{ni}$, $x_{nj} (i,j=1,2,3)$ are the 3D position vector of the 
particle with respect to the ELG center. 
For this method, the weight $1/r_n^2$ is used, but the following results we show remain almost unchanged even if we do not use this radial weight, as long as the summation is restricted to member particles of each ELG.

The ellipticity of a galaxy is (assuming the $x_3$-axis as the LOS direction):
\begin{equation}
    \epsilon_1\equiv\frac{I_{11}-I_{22}}{I_{11}+I_{22}}, \epsilon_2 \equiv\frac{2I_{12}}{I_{11}+I_{22}}.
\label{eq:e_inertia}
\end{equation}
In the following we use either of Eq.~(\ref{eq:aperture_Iij}) or (\ref{eq:reduced_Iij}) for the inertia tensor.
The column ``$\sigma_\epsilon$'' in Table~\ref{tab_elg} gives the intrinsic rms ellipticities for the new method (Eq.~\ref{eq:aperture_Iij}), showing that the new method gives a larger 
$\sigma_\epsilon\sim 0.4$ than that of the usual method, 
$\sigma_\epsilon \sim 0.3$ as shown in \citet{2021JCAP...03..030S}.

The IA power spectrum between matter density field $\delta_m$ and $E$-mode shear field $\gamma_E$ is estimated following the method in \citet{2021MNRAS.501..833K}:
\begin{equation}
    \langle\gamma_E(\bm{k})\delta_m(\bm{k'})\rangle \equiv 
    (2\pi)^3\delta_D(\bm{k}+\bm{k'})P_{\delta E}(\bm{k}),
\label{eq:ps_me}
\end{equation}
where $\gamma_E(\bm{k})=\gamma_{1}(\bm{k}) \cos{2\phi_{\bm{k}}}+\gamma_{2}(\bm{k}) \sin{2\phi_{\bm{k}}}$ is the $E$-mode decomposition of galaxy shear in Fourier space and $\gamma_{1,2}=\epsilon_{1,2}/(2{\cal R})$ ($\mathcal{R} \equiv 1-\langle \epsilon_i^2\rangle$ is the responsivity as defined in \citealt{2002AJ....123..583B}). The non-linear alignment model \citep{2011JCAP...05..010B} predicts 
\begin{equation}
P_{\delta E}(k,\mu)=-A_{\rm IA}C_1\rho_{\rm cr0}\frac{\Omega_{\rm m}}{D(z)}(1-\mu^2)P_{\delta\delta}(k,z),
\label{eq:A_IA_dE}
\end{equation}
where $P_{\delta\delta}(k,z)$ is the non-linear matter power spectrum at redshift $z$, $D(z)$ is the growth rate, and $C_1\rho_{\rm cr0}=0.0134$ for convention \citep{Joachimi2011}. The dimension-less coefficient $A_{\rm IA}$ is an indicator of the IA strength \citep{2021JCAP...03..030S}. 

\section{Results} 
\label{sec_res}

\begin{figure}
\centering
 \includegraphics[width=1.\linewidth]{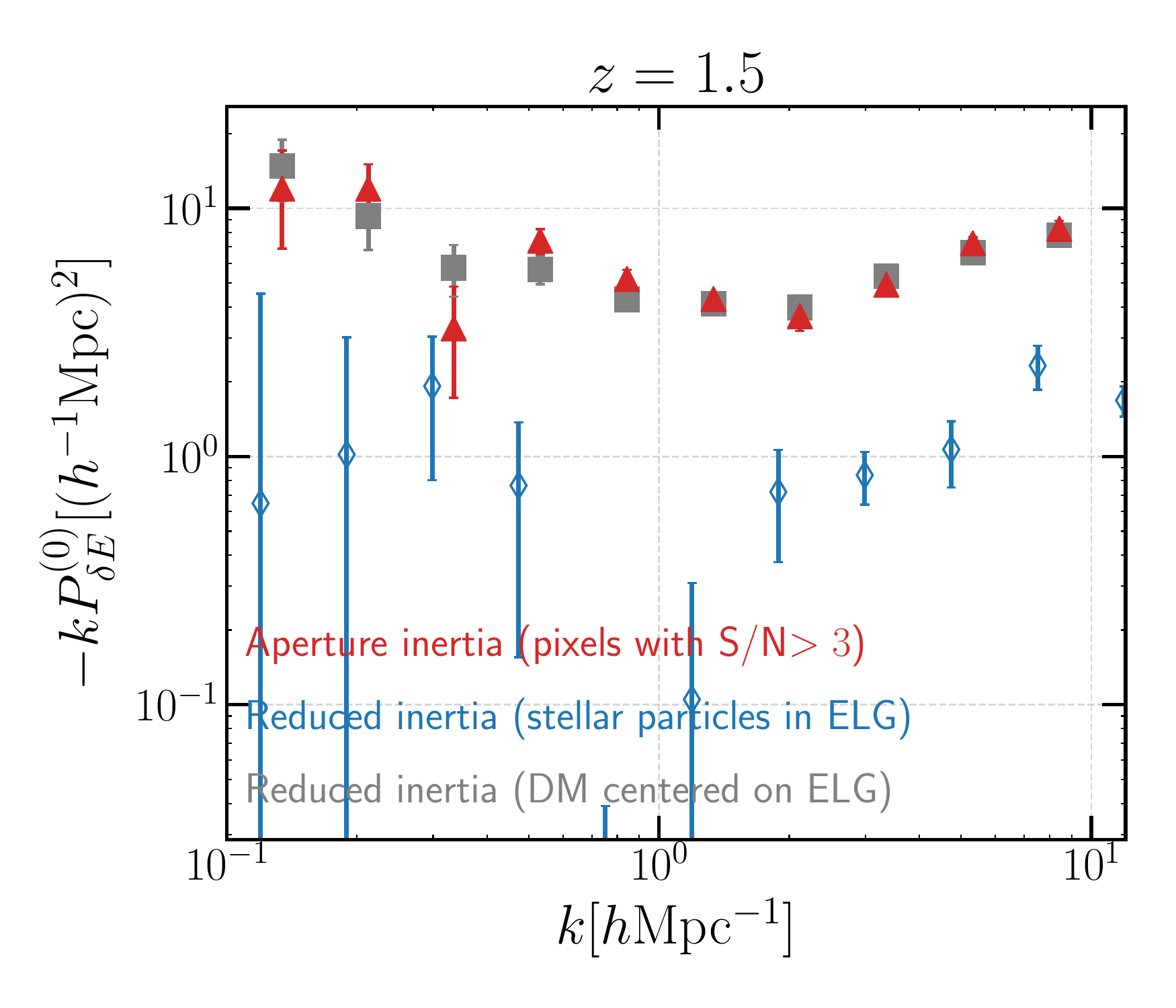}
\caption{\label{fig_PS_IA} The monopole moment of the cross-power spectrum between matter and the galaxy $E$-mode shape, $P^{(0)}_{\delta E}(k)$, for ELGs at $z=1.5$.
The red triangles show the result when using the aperture shape estimator (Eq.~\ref{eq:aperture_Iij}). For comparison the open diamonds show the result when using the standard method of shape estimator (Eq.~\ref{eq:reduced_Iij})
for the same sample of EGLs, and the squares show the result using the standard method for the host halos using DM particles centered on each ELG.}
\end{figure}

\begin{figure}
\centering
 \includegraphics[width=.95\linewidth]{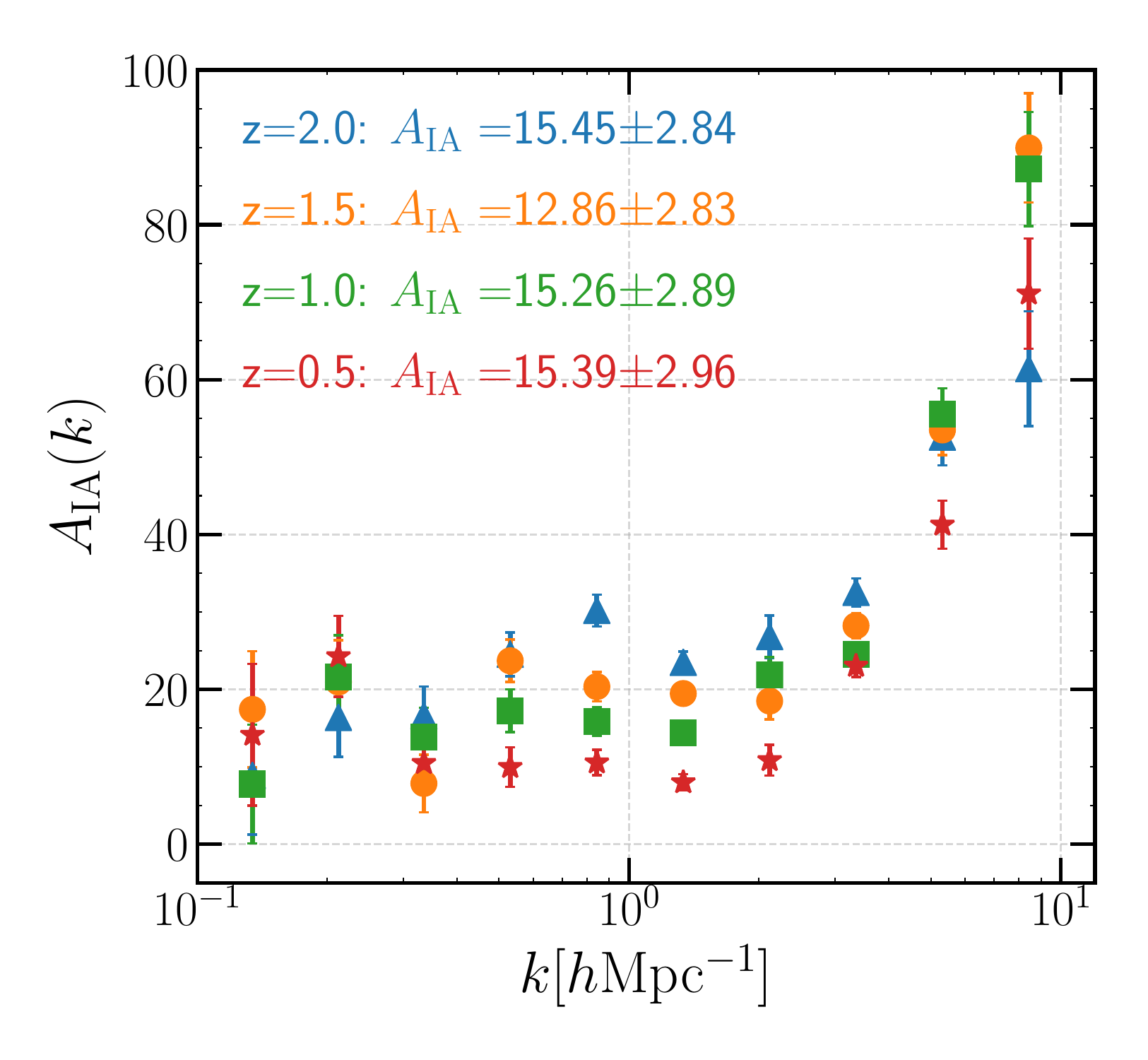}
\caption{\label{fig_AIA_zevol} 
The IA strength, characterized by $A_{\rm IA}(k)\propto P^{(0)}_{\delta E}/P_{\delta\delta}$, for the ELG samples of the fixed number density at different redshifts. The $A_{\rm IA}$ values in the legend are the best-fit linear IA coefficient, obtained from the data points with $k<0.4h{\rm Mpc^{-1}}$ (see text for details).}
\end{figure}

In this section we show the main results of this paper. Fig.~\ref{fig_PS_IA} shows that our new estimator of ELG shape, defined by Eq.~(\ref{eq:aperture_Iij}), allows for a clear detection of the monopole moment of $P_{\delta E}^{(0)}(k)$ at $z=1.5$, while the conventional shape method gives only an upper limit on the cross-power spectrum in low $k$ bins. To be more quantitative, the new method gives more than a ten-fold boost in the $P^{(0)}_{\delta E}$ amplitude over the range of $k$ bins we consider. Recalling that the TNG300 simulation has a small volume of $\sim 0.0086~(h^{-1}{\rm Gpc})^3$, this result means that upcoming galaxy surveys covering more than $1\,(h^{-1}{\rm Gpc})^3$ volume enables a significant detection of the IA signal. The IA signal in smaller $k$ bins contains cleaner cosmological information, and a fitting of the model (Eq.~\ref{eq:A_IA_dE}) with the measured power spectrum over the three lowest $k$-bins (up to $k\simeq 0.4\,h{\rm Mpc}^{-1}$) 
gives $A_{\rm IA}=12.86\pm 2.83$, 4.5$\sigma$ detection, while $A_{\rm IA}$ for the standard method is consistent with a null detection at $2\sigma$ level ($A_{\rm IA}=3.2\pm 2.0$). 
Is this new IA estimator optimal? To address this question, Fig.~\ref{fig_PS_IA} also shows the IA power spectrum for DM halos hosting ELGs, where we use the parent halos even for satellite ELGs (the member DM particles) to characterize the halo shapes centered on each ELG.
The DM halo gives $A_{\rm IA}=15.7\pm 2.4$, which is very similar to the IA signal of ELGs. The good agreement between the flux based aperture inertia tensor and the DM halo particle based inertia tensor suggests that the light distribution follows the matter distribution, which is supported by the good correlations (although with scatters) between the ellipticities calculated using light and matter distributions as shown in Appendix~\ref{sec_ellipticities}. This is also consistent with the results shown in \citet{2021arXiv210505914S} and \citet{2021MNRAS.504.4649O}, where they show the baryons trace the matter distribution well using DES lensing profile and IllustrisTNG hydro-simulation separately. 
Also, the stronger IA signal with our aperture based inertia tensor is in consistent with the picture that the outer region of galaxies/satellites in galaxy groups are more aligned with the large scale tidal field as revealed by previous studies \citep{2016MNRAS.457.2301S,2017MNRAS.467.4131V}.

Table~\ref{tab_elg} summarizes the IA signal for ELGs at different redshifts, $z=0.5, 1.0, 1.5$ and $2.0$. Note that all ELG samples have the fixed number density of $10^{-4}\,(h^{-1}{\rm Mpc})^3$. 
The ELG samples at all the redshifts give a clear detection of $A_{\rm IA}$. Fig.~\ref{fig_AIA_zevol} shows the ratio of the IA power spectrum to the matter power spectrum. 
The figure shows that the IA signals are detected over the range of $k$ bins, with very weak redshift dependence. 
The redshift evolution of $A_{\rm IA}$ depends on the sample selection and redshift, as shown in Fig.~$6$ of \cite{2021MNRAS.501..833K}.
In our previous work of \citet{2021JCAP...03..030S}, we found that $A_{\rm IA}$ shows very weak redshift dependence for the galaxy samples of a fixed stellar mass range across $z=0.3$ to $z=2$. The mean stellar mass varies within $\sim 0.3$ dex from $z=0.5$ to $z=2$ for the ELGs, as listed in Table~\ref{tab_elg}. The weak redshift dependence in the ELG IA signals is thus consistent with our previous studies. 
In addition the ratio displays very weak $k$-dependence up to $k\sim 1\hmpci$, which is in agreement with the prediction of non-linear alignment model.

In Appendix~\ref{sec_aperture_size} we also show how the results change with varying aperture sizes
and the $S/N$ cuts in the pixels that are needed to define the aperture based shapes of ELGs. Figs.~\ref{fig_AIA_zevol} and~\ref{fig_xSN} show that the findings we described above hold for these different definitions of the ELG shapes.

\section{Discussion and Summary}
\label{sec_con}

In this paper we proposed an aperture-based estimator to characterize shapes of ELGs (more exactly star-forming galaxies) for 
extracting the IA signals of ELGs. We applied the method to star-forming galaxies simulated in IllustrisTNG, one of the state-of-the-art cosmological hydrodynamical simulations, and showed that the method allows for a significant detection of the IA effect even from the small simulation volume.
This method gives about ten-fold boost in the IA amplitude  compared to that of the conventional method. We also found a significant detection of the IA signals, with almost similar amplitude and signal-to-noise ratios, for all the ELG samples over the wide range of redshifts up to $z=2$. 
This is quite encouraging because the new method opens up an opportunity to study the IA signals of star forming galaxies over redshifts where the cosmic star formation activity is violent, and the measured IA signals can be used to probe cosmology and physics of galaxy formation. This method is relevant for upcoming imaging and spectroscopic galaxy surveys: Subaru HSC, PFS, DESI, LSST, Euclid and Roman Space Telescope.

In order to characterize the shape of each ELG region, we need to properly estimate the background noise in each field and then use the pixels that are greater than a certain threshold ($S/N>3$ used in this paper). Defining the uniform background noise over the entire survey region is not so obvious, as it depends on the depth and sky brightness of each pointing in each field. As long as the background estimation is random between different fields, this does not cause any systematic effect in the IA measurement. If the background estimation varies with different fields in a correlated way with large-scale structures for some reason (e.g. contribution from light of galaxies in each field), it would cause the systematic effect. For the similar reason, any projection effect of foreground/background (physically unassociated) galaxies in each ELG region relevant for the shape estimation causes only statistical noise in the IA measurement. A practical application of the method to actual data would be quite valuable and will be our future work.

\acknowledgements
We thank R. K. Sheth and Elisa Chisari for enlightening discussion/comments on this work. 
J. Shi thanks Junyao Li for useful discussions on observational effects.
This work was supported in part by World Premier International Research Center Initiative (WPI Initiative), MEXT, Japan, and JSPS KAKENHI Grant Numbers JP18H04350, JP18H04358, JP19H00677, JP20J22055, JP20H05850, and JP20H05855.
KO is supported by JSPS Overseas Research Fellowships.
TK is supported by JSPS Research Fellowship for Young Scientists.

\bibliographystyle{mn2e_new}
\bibliography{ref}

\appendix
\counterwithin{figure}{section}

\section{Ellipticities of ELGs}
\label{sec_ellipticities}

\begin{figure}
\centering
 \includegraphics[width=0.9\linewidth]{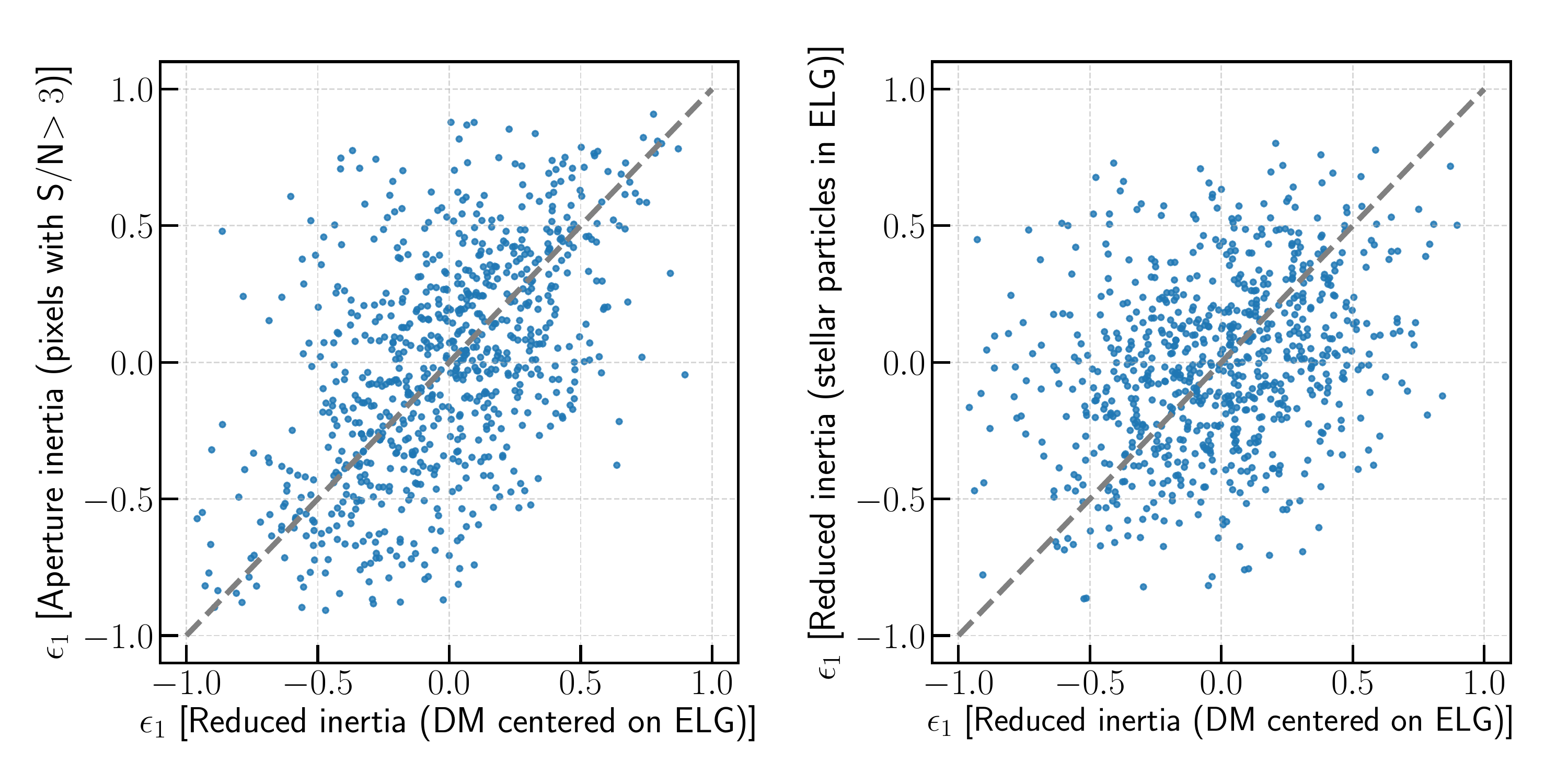}
\caption{\label{fig_e} {\it Left panel}: A comparison of 
the ellipticities for ELGs at $z=1.5$ calculated using the aperture shape estimator (Eq.~\ref{eq:aperture_Iij})
with the standard method of shape estimator  (Eq.~\ref{eq:reduced_Iij}) for the host halos using DM particles centered on each ELG on individual ELG basis.  
{\it Right}: The simplar plot, but the comparison of the 
DM halo shape, the same as in the left panel, with the ELG shapes obtained by applying the standard method of shape estimator to the stellar particles of each ELG.
The Spearman's rank correlation coefficients for the left and right panels are $0.524$ and $0.289$ separately.
The gray dashed lines correspond to perfect correlations for reference.}
\end{figure}

In Fig.~\ref{fig_e}, we show the ellipticity for each ELG at $z=1.5$ calculated using the aperture shape estimator developed in the this work versus the ellipticity calcualted using the DM particles within the host halo centering on the the ELG. The ellipticity calculated using the standard method with stellar particles within the galaxy is also shown for comparison. The figure clearly shows that the new method of ELG shapes gives a stronger correlation with the DM halo shapes than the conventional method does. Although large scatters exist in this one-to-one ellipticity correlation (see Appendix D of \citealt{2021MNRAS.501..833K} for the related discussion), the nice thing about the power spectrum method is that it allows us to extract a correlated signal between shapes of different ELGs, where the intrinsic shapes act as statistical errors.

\section{IA power spectrum with varying aperture sizes
and $S/N$ cuts}
\label{sec_aperture_size}

\begin{figure}
\centering
 \includegraphics[width=.9\linewidth]{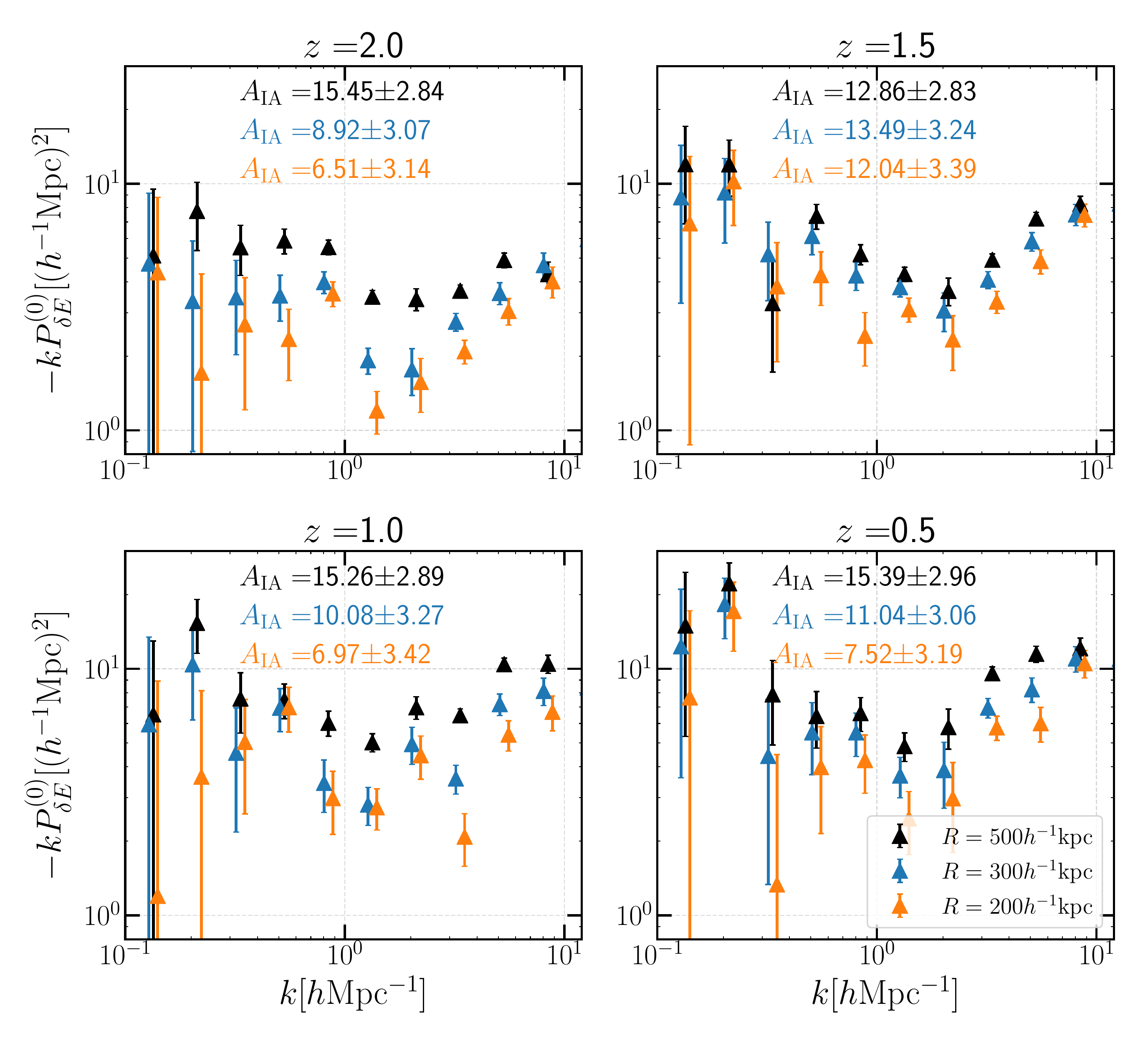}
\caption{\label{fig_aperture_size} The IA power spectrum of ELGs measured with varying aperture radii that are 
used to define the ELG shapes (Eq.~\ref{eq:aperture_Iij}). 
The black, blue, and orange triangles with errorbars are the results with radius of $500\kpc$, $300\kpc$, and $200\kpc$. The IA strength characterized by $A_{\rm IA }$ are also shown in the figure.}
\end{figure}

In Fig.~\ref{fig_aperture_size}, we study the dependence of the IA strength on the aperture size choices. The IA strength is stronger and has a higher $S/N$ ratio with aperture radii increasing from $200\kpc$ to $500\kpc$. The IA strength with $300\kpc$ approaches the one with $500\kpc$. Such dependence on aperture size is in agreement with the point made in Fig.~\ref{fig_PS_IA}, where the IA signal is weak/non-detectable when we use the reduced inertia tensor based on stellar particles and the signal gets stronger and clearer when we include and give more weight to the outer region within the host halo.

\begin{figure}
\centering
 \includegraphics[width=.6\linewidth]{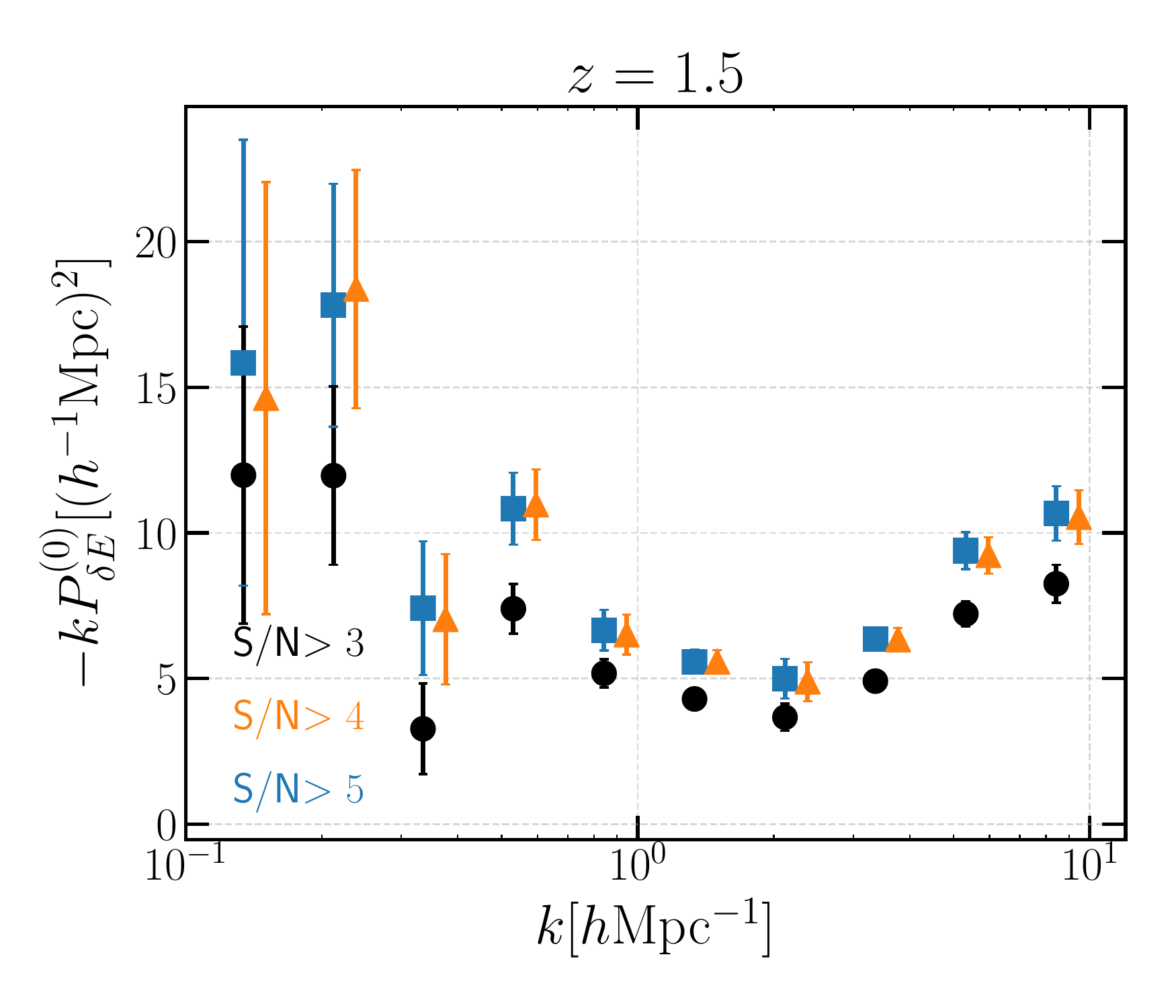}
\caption{\label{fig_xSN} The IA power spectrum of ELGs measured with varying the $S/N$ cut used for the ELG shape definition (Eq.~\ref{eq:aperture_Iij}).}
\end{figure}

Fig.~\ref{fig_xSN} shows the results using the different $S/N$ cuts of pixels that are used to define the ELG shapes (see Eq.~\ref{eq:aperture_Iij} and Fig.~\ref{fig_gal_image}). It is clear that the IA signals remain for the different choices of the $S/N$ cuts.

\end{document}